\documentclass[twocolumn,showpacs,reprint,superscriptaddress,amsmath,amssymb]{revtex4-1}
\usepackage{hyperref}
\usepackage{graphicx}% Include figurefiles

\newcommand{\onebody}[3]{\ensuremath{\langle{#1}|{#2}|{#3}\rangle}}

\newcommand{\bvec}[1]{\ensuremath{\boldsymbol{#1}}}

\newcommand{\tgenm}[1]{\ensuremath{\boldsymbol{\mathcal{#1}}}}

\newcommand{\matnot}[1]{\ensuremath{\mathbf{#1}}}

\newcommand{\op}[1]{\ensuremath{\hat{#1}}}

\newcommand{\lat}[1]{\ensuremath{\mathfrak{#1}}}

\SetMathAlphabet{\mathcal}{normal}{OMS}{mdbch}{m}{n}
\begin{document}

\title{A doubly-periodic structure for the study of inhomogeneous bulk fermion matter with spatial localizations}

\date{\today}
\author{Klaas Vantournhout} \email{k.vantournhout@gsi.de}
\affiliation{GSI Helmholtzzentrum f\"ur Schwerionenforschung GmbH,
  Planckstra\ss e 1, 64291 Darmstadt, Germany}
\affiliation{Department of Physics and Astronomy, Ghent University, Proeftuinstraat 86, 9000 Gent, Belgium}
\author{Natalie Jachowicz}
% \email{natalie.jachowicz@ugent.be}
\affiliation{Department of Physics and Astronomy, Ghent University,
  Proeftuinstraat 86, 9000 Gent, Belgium}
\author{Jan Ryckebusch}
% \email{jan.ryckebusch@ugent.be}
\affiliation{Department of Physics and Astronomy, Ghent University,
  Proeftuinstraat 86, 9000 Gent, Belgium}

\begin{abstract}
We present a method that offers perspectives to perform fully antisymmetrized simulations for inhomogeneous bulk fermion matter. The technique bears resemblance to classical periodic boundary conditions, using localized single-particle states. Such localized states are an ideal tool to discuss phenomena where spatial localization plays an important role. The antisymmetrisation is obtained introducing a doubly-periodic structure in the many-body fermion wave functions. This results in circulant matrices for the evaluation of expectation values, leading to a computationally tractable formalism to study fully antisymmetrized bulk fermion matter. We show that the proposed technique is able to reproduce essential fermion features in an elegant and computationally advantageous manner.

\end{abstract}

\maketitle

Bulk fermion systems are ubiquitous in nature. Mostly studied as periodic homogeneous structures, they often exhibit spatial localizations resulting from impurities, random external fields or local minima in the overall potential \cite{Altshuler1980,Sanpera2004,Watanabe2005}. The crust of a neutron star is an example of such a system. It is build up from protons, neutrons and electrons governed by short-range nuclear attraction and long-range Coulomb repulsion. As a result of these interactions and the subnuclear densities of the neutron star's crust, spatial localizations play a crucial role. In the lower-density regions of the crust, the crustal matter organizes itself in a Coulomb lattice, while in the higher-density regions (i.e.~the crust-core interface), a subtle interplay between those interactions lead to complex structures dubbed ``nuclear pastas.''  With a preponderance of low-energy excitation levels, these shapes are susceptible to low-energy dynamics stemming from external probes or temperature changes. Molecular dynamics techniques are appropriate to study such systems. Of the many existing techniques \cite{Allen1991, Aichelin1986, Maruyama1996, Feldmeier2000, Ono2004}, antisymmetrized molecular dynamics (AMD) and fermionic molecular dynamics (FMD) are the ones that allow a full quantum treatment of a fermion system, using an antisymmetrized set of localized states. However, a formalism to employ AMD or FMD on bulk systems with full antisymmetrization is still missing.

In this paper, we introduce a technique for simulating bulk fermion systems that is based on a doubly-periodic structure in the many-body wave function. The method makes use of localized nonorthogonal single-particle states and allows one to address spatial localizations. We demonstrate how a complete antisymmetrization of localized states under periodic boundary conditions can be achieved in a computationally attractive fashion. Furthermore, the presented technique is directly applicable to AMD and FMD.\\ 

When investigating the properties of bulk matter by means of large simulation volumes, the evaluation of expectation values becomes cumbersome. Moreover, surface effects may influence the results when a large fraction of the constituents lies on the surface of the simulation volume. A time-honored method is to introduce a periodic structure in the simulation. When studying bulk matter using a periodic structure, it is crucial to make sure that the studied properties of the small but infinitely repeated periodic system and the macroscopic system which it represents, are the same. As long as the correlation volume of the interactions does not exceed the simulation volume, the imposed periodicity works fine. However, serious problems arise in the presence of long-range correlations such as for example those induced by Coulomb interactions or by the Pauli exclusion principle.

When studying fermion systems in a mean-field approach, the many-body state is often introduced as an antisymmetrized product of single-particle states \cite{Slater1929}. For the description of bulk matter, the single-particle states of Slater determinants generally fulfill certain boundary conditions. Within a periodic structure, the full Hilbert space of these single-particle states is spanned by Bloch-Floquet states \cite{Floquet1883, Bloch1929, Ashcroft1976}. These are, however, tedious to evaluate and often a subset of the Hilbert space is used. This subset is generally referred to as ``periodic boundary conditions'' or the more general ``twist-boundary conditions''. The undesired finite-sized shell effects stemming from such a restricted set are much reduced when averaging over the twist angle. This technique is generally referred to as ``phase-randomization'' or ``twist-averaged boundary conditions'' \cite{TinkaGammel1993,Lin2001}.

To describe, however, spatial localizations with delocalized states, a large configuration space is required. To circumvent this problem Wannier states can be used as single-particle states. Wannier functions span the full Hilbert space of the periodic system and represent localized states \cite{Wannier1937, Blount1962, Ashcroft1976}.

All previously proposed states fulfill periodic properties by definition. In this paper, we propose to represent the inhomogeneous bulk system using nonorthogonal localized states $\phi(\bvec r)$, independent of any periodic framework, and embed them in a periodic-boundary framework. As the Pauli exclusion principle introduces long-range many-body correlations, it poses a major challenge to the study of bulk fermion systems with nonorthogonal localized states. It is indispensable to fully antisymmetrize the many-body wave function representing the system.

The properties of a Fermi system are studied by evaluating various $N$-body operators. For a many-body fermion wave-packet $|\Phi\rangle$, denoted as
%---------------------------------------------------------------------
\begin{equation}
|\Phi\rangle = \op{A}|\phi_1\rangle\otimes\cdots\otimes|\phi_A\rangle,
\end{equation}
%---------------------------------------------------------------------
the expectation values of one- and two-body operators are calculated as
%---------------------------------------------------------------------
\begin{subequations}\label{eq:finite:operators}
\begin{align}
\mathcal{B}_I &=
\sum_{pq=1}^A\onebody{\phi_p}{\op{B}_{I}}{\phi_q}\matnot{o}_{qp},\label{eq:finite:operators:1B}\\
\mathcal{B}_{II}&=
\frac{1}{2}\sum_{pqrs=1}^A \onebody{\phi_p\phi_r}{\op{B}_{II}}{\phi_q\phi_s}
(\matnot{o}_{qp}\matnot{o}_{sr} -
\matnot{o}_{qr}\matnot{o}_{sp}).\label{eq:finite:operators:2B}
\end{align}
\end{subequations}
%---------------------------------------------------------------------
Here, the matrix $\matnot o$ represents the inverse of the overlap matrix $\matnot n$ with $\matnot n_{pq} = \langle \phi_p|\phi_q\rangle$. Because of the determinant structure of $|\Phi\rangle$, the expectation values can be written as traces using the inverse overlap matrix $\matnot o$. Thereby, the matrices $\matnot n$ and $\matnot o$ play a fundamental role because they carry all the information about the fermion statistics of the system under study. As their eigenvalues can cover many orders of magnitudes, it is paramount to calculate these matrices as accurately as possible (i.e.~analytically). Due to the long-range character of the Pauli correlations, this becomes a tedious task for bulk fermion matter. The dimension of the matrices usually impedes simulations with a large number of particles, however desirable these may be for bulk fermion systems.

\begin{figure}[h!b]
  \begin{center}
    \includegraphics[width=0.4\textwidth]{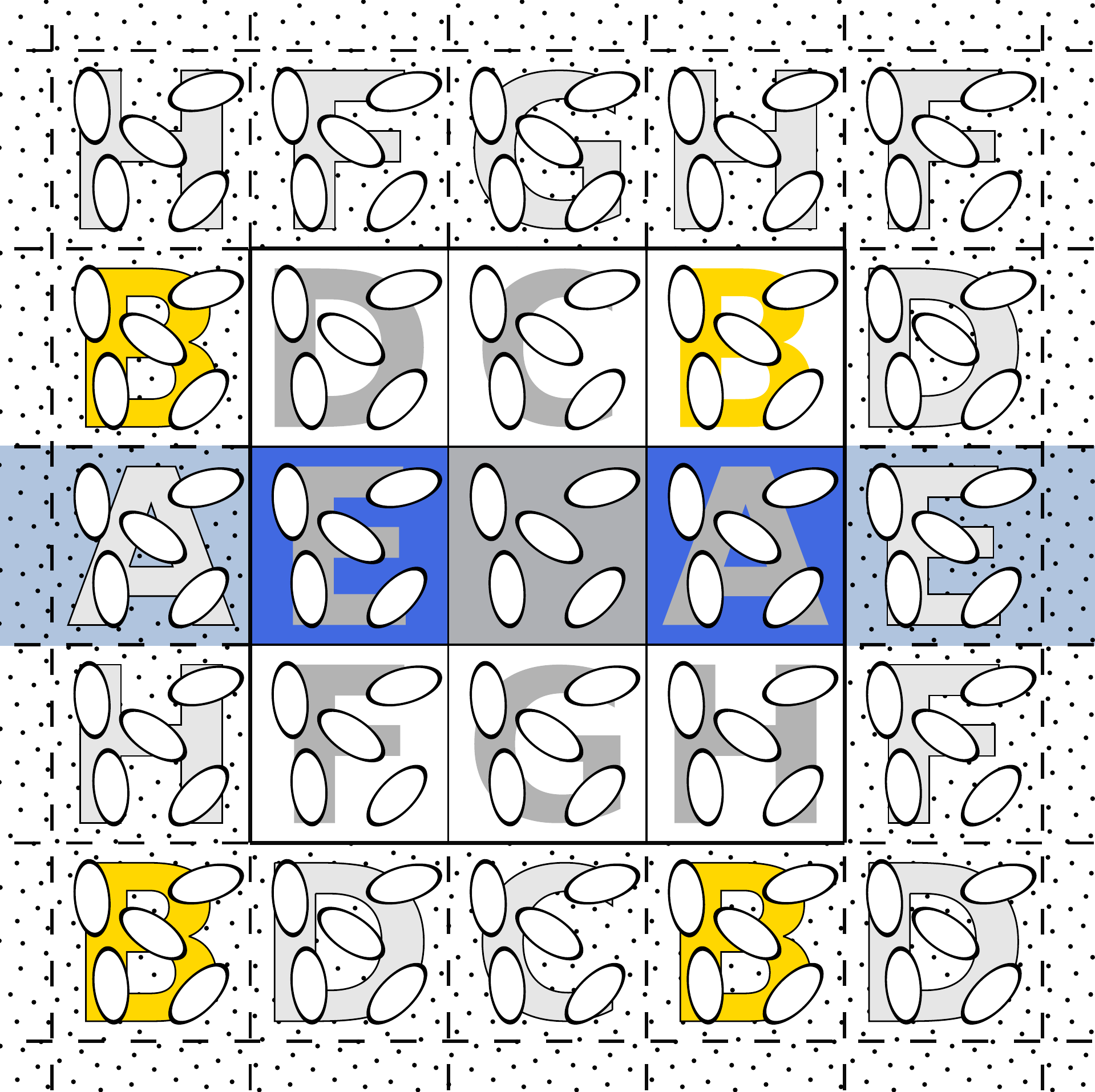}
\caption{(Color online) Simulation volume (full-bordered cells) for a two-dimensional periodic system of first order, created by a unit-cell (gray-shaded cell) and eight identical copies (A to H) with box length $\ell$, symmetrically placed on the lattice sites of a two-dimensional square lattice. In addition, the area outside the actual simulation volume (dotted area) is generated by the particles' wave functions that have a periodicity $3\ell$ in both directions. Therefore, the constituents of boxes with identical labels are the same. Because of the double periodicity, the constituents in cells on the border of the simulation volume not only feel the influence of the particles in the neighboring cells but also of all other particles. For instance, the particles in cell A feel the effect of the particles in the central cell and in cells B, C, G and H directly, while the periodicity of the wave functions introduces indirect interactions with the particles in all emanations of cells D, E and F. Hence, the double periodicity mimics an infinite simulation volume in an effective way. The method is iterative: while the two-dimensional simulation volume can be created by a single unit-cell, it can also be seen as the periodic extension of a one-dimensional periodic simulation volume (the blue-shaded row in the figure) created by the unit cell in the center.}
    \label{fig:TAPBC}
  \end{center}
\end{figure}

Two general approaches exist for creating bulk systems\,: positioning the single-particle states periodically or using periodic single-particle wave functions. While the first technique leads to infinite matrices, the second results in a limited antisymmetrization. We propose a technique which overcomes these problems by introducing a doubly periodic structure in the description of bulk fermionic matter. As can be seen in Fig.~\ref{fig:TAPBC}, both the spatial positioning of the simulation volume and the single-particle wave packets are made periodic.
%---------------------------------------------------------------------
%---------------------------------------------------------------------
A number of particles are placed in unit cells on a lattice \lat B, which tessellate space perfectly. Each cell, containing $A$ particles, can be identified by a lattice vector $\bvec R=n_1\bvec a_1 + n_2\bvec a_2 + n_3\bvec a_3$ with integer $n_j$. A simulation of order $m$ consists of the cells on the lattice vectors defined by the condition $n_j\in\{-m,\ldots,m\}$, and contains $(2m+1)^3$ identical copies. To eliminate the surface effects, the single-particle states are subjected to the Born-von K\'arm\'an boundary conditions \cite{Born1912, Ashcroft1976} with periodicity $(2m+1)\bvec a_j$. After these manipulations, the many-body fermion Slater determinant is written as
%---------------------------------------------------------------------
\begin{equation}\label{eq:tapbc:W}
  |\Phi_m\rangle = \op{A}\bigotimes_{\bvec R\in\{\lat B\}_m} \op{T}(\bvec R)
  \left\{|\phi_{m,1}\rangle\otimes\cdots\otimes|\phi_{m,A}\rangle\right\},
\end{equation}
%---------------------------------------------------------------------
where $\{\lat B\}_m$ is the finite set of lattice vectors, $\op{T}(\bvec R)$ the translation operator over $\bvec R$, and
%---------------------------------------------------------------------
\begin{equation}\label{eq:tapbc:periodic:wq}
  |\phi_{m,p}\rangle = \sum_{\bvec R\in\lat B} 
  \op T((2m+1)\bvec R) |\phi_p\rangle.
\end{equation}
%---------------------------------------------------------------------
The trial state $|\Phi_m\rangle$ can be understood as an infinite system with a truncated range of the antisymmetry operator. Alternatively, $|\Phi_m\rangle$ can also be interpreted as a finite system with limited periodicity mapped on a toroidal structure. In case of the example given in Fig.~\ref{fig:TAPBC}, the original 9 cells (A through F and the central gray cell) are mapped onto a torus. The dotted regions are then obtained by unfolding the torus while keeping the toroidal boundary conditions intact [see Eq.~\eqref{eq:tapbc:periodic:wq}].

Due to the doubly-periodic structure of the trial state $|\Phi_m\rangle$, the fundamental overlap matrix and its inverse exhibit a peculiar nested block-circulant structure. This circulant structure can be exploited to map the description of the complete trial state onto a set of equations related to only one unit cell. For the overlap matrix of a one-dimensional system, denoted as $\matnot N$, the circulant structure is given by
%---------------------------------------------------------------------
\begin{equation}\label{eq:tapbc:circulant}
\includegraphics[width=0.4\textwidth]{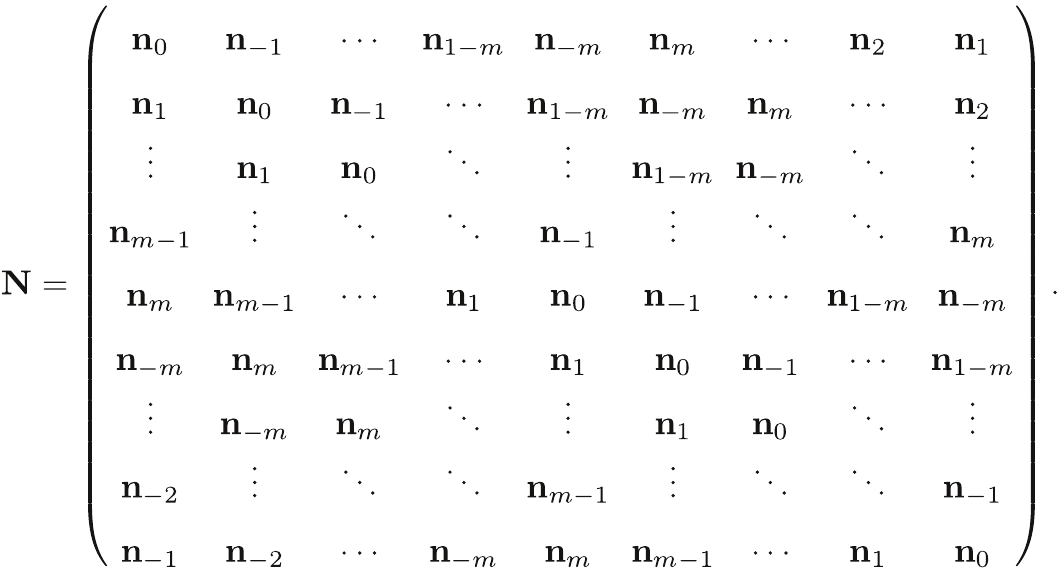}
\end{equation}
%---------------------------------------------------------------------
The blocks $\matnot n_{k}$ are $A\times A$ matrices with elements $\matnot n_{pq,k} = \langle \phi_{m,p}|\op{T}(-k\bvec a_1)|\phi_{m,q}\rangle$. For a two-dimensional periodic system, these blocks are $(2m+1)A\times(2m+1)A$ one-dimensional overlap matrices. This is illustrated in Fig.~\ref{fig:TAPBC}. This modular structure can be extended to higher dimensions, resulting in overlap matrices with a nested block-circulant structure. As a result of the translation invariance of the system, each $A\times A$ block of the overlap matrix can be identified unambiguously with the lattice vector $\bvec R$ that connects the two cells of the bra and ket states. The blocks can be evaluated as $\matnot n_{pq,\bvec R} = \langle \phi_{m,p}|\op{T}(-\bvec R)|\phi_{m,q}\rangle$.

Genuine antisymmetrization of an infinite system representing bulk matter cannot be achieved in a restricted simulation volume. It requires that the equations be evaluated for $m\rightarrow\infty$. Under those circumstances, the wave packets $|\phi_{\infty,p}\rangle$ have infinite periods and match the original localized single-particle states $|\phi_p\rangle$. The inverse overlap matrix can be obtained through the following scheme \cite{vantournhout2010}\,:
%---------------------------------------------------------------------
%\begin{subequations}
\begin{gather*}
\tgenm N(\bvec k) = \sum_{\bvec R\in\lat B} \matnot n_{\bvec R}
  \,e^{-i\bvec k\cdot\bvec R},\qquad
\tgenm O(\bvec k) = \tgenm
  N(\bvec k)^{-1}, \\
\matnot o_{\bvec R} = \frac{1}{V_{BZ}}\int_{BZ}\tgenm O(\bvec
  k)\,e^{i\bvec k\cdot\bvec R}\,d^3\bvec k,
\end{gather*}
%\end{subequations}
%---------------------------------------------------------------------
where $BZ$ represents the first Brillouin zone of the lattice $\lat B$ and $\bvec k$ is a vector in this volume. The strength of the proposed formalism reveals itself upon evaluating the operators in reciprocal space. Normally, the expectation values of the infinite fermion system would be calculated by means of Eqs.~\eqref{eq:finite:operators}, resulting in infinite sums over the block structure\,:
%---------------------------------------------------------------------
\[
\mathcal{B}_{\rho,I} =  \sum_{\bvec R\in\lat B}  \sum_{pq=1}^A \onebody{\phi_p}{\op{B}_{I}\op{T}(\bvec R)}{\phi_q}\matnot{o}_{\bvec R,qp}.
\]
%---------------------------------------------------------------------
 These sums, however, translate into integrals over the first Brillouin zone which are computationally straightforward to evaluate. The expectation value of a one-body operator per unit-cell volume can then be computed as
%---------------------------------------------------------------------
\begin{subequations}\label{eq:pbc:operators:bloch}
  \begin{align}
    \mathcal{B}_{\rho,I} &= \frac{1}{V_{BZ}}\int_{BZ} \sum_{pq=1}^A\tgenm
    B_{I,pq}(\bvec k)\tgenm O_{qp}(\bvec k)\,d^3\bvec k ,\\
    \tgenm B_{I,pq}(\bvec k) &= \sum_{\bvec R\in\lat B}
    \onebody{\phi_p}{\op{B}_{I}\op{T}(\bvec R)}{\phi_q}
    \,e^{i\bvec k\cdot\bvec R}.
  \end{align}
\end{subequations}
%---------------------------------------------------------------------
Here we assume that operator $\op{B}_I$ commutes with the translation operator $\op{T}(\bvec R)$. The two-body operator has an analogous structure.

From a computational perspective, a proper choice of the localized single-particle states helps speeding-up the computations. Gaussian wave packets come with the interesting feature that most matrix-elements can be evaluated analytically. Furthermore, the computational order for the calculation of expectation values is the same for bulk fermion systems as for finite-sized systems ($N^4$ for two-body interactions), even though the former represent a system with an infinite number of particles. On the other hand, the equations for the bulk systems involve an extra integration over the Brillouin zone of the imposed lattice, increasing the computational effort. However, it has been shown that integrals over the Brillouin zone can be performed using only a limited number of specific $\bvec k$ points \cite{Baldereschi1973,Monkhorst1976}.\\

%---------------------------------------------------------------------
\begin{figure}[h!t]
\begin{center}
\includegraphics[width=0.5\textwidth]{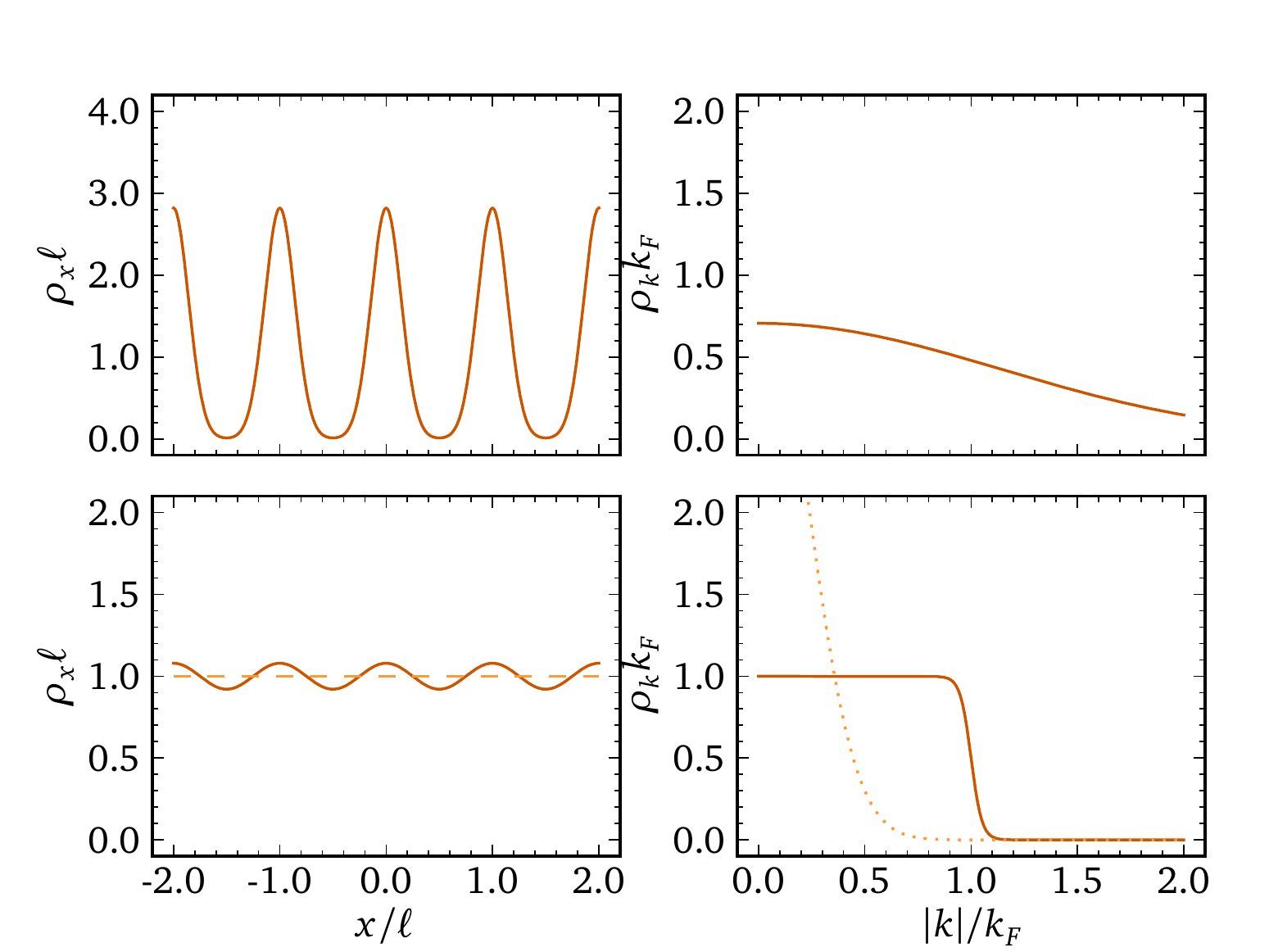}
\caption{(Color online) Spatial density (left panels) and momentum density (right panels) of equidistant Gaussian wave packets with one-dimensional periodicity. One Gaussian with variance $a$ per unit cell of size $\ell$. For small overlap (top\,: $\sqrt{a}/\ell=0.2$) the system behaves like one of distinguishable particles while for large overlap (bottom\,: $\sqrt{a}/\ell=1$) the fermionic behavior becomes manifest. For comparison, the spatial distribution of a uniform Fermi gas ($\sqrt{a}/\ell\rightarrow\infty$) is presented (dashed line). The momentum distribution of distinguishable particles is also presented (dotted line). The value of $k_F$ is given by $\pi/\ell$.}\label{fig:1ddist}
\end{center}
\end{figure}
\begin{figure}[b]
\begin{center}
\includegraphics[width=0.5\textwidth]{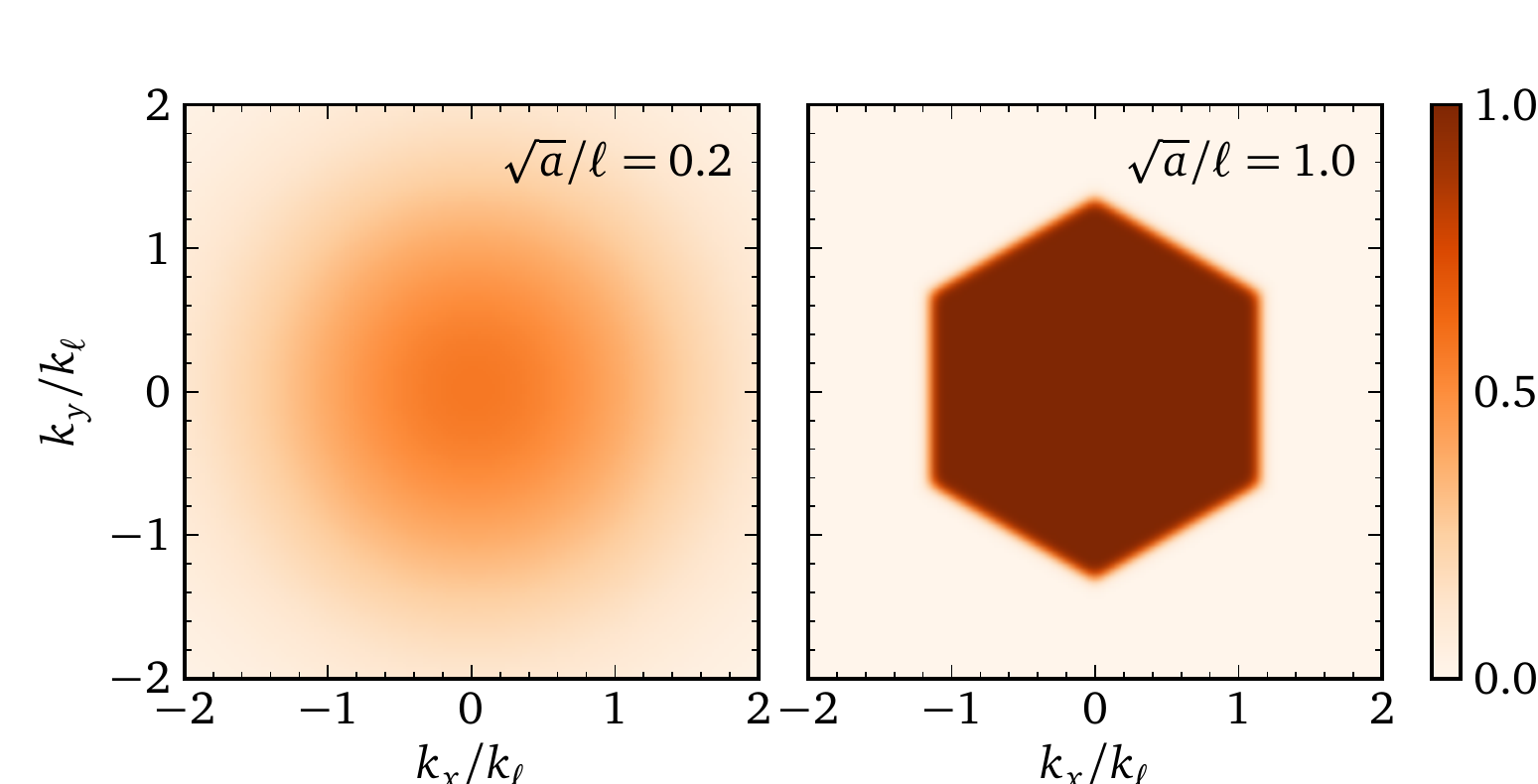}
\caption{(Color online) The momentum distribution of a single fermion under hexagonal
  boundary conditions. With increasing overlap the shape of the
  momentum distribution evolves toward that of the first Brillouin
  zone. The value of $k_\ell$ is given by $\pi/\ell$.}\label{fig:2dhexdist}
\end{center}
\end{figure}

%---------------------------------------------------------------------
In the following, we will show that the proposed technique reproduces the features intrinsic to the fermionic behavior of the system. The single-particle states are Gaussian wave packets of the form $\langle\bvec x|a\bvec b\rangle = \exp\{-(\bvec x - \bvec b)^2/(2a)\}$ where the complex vector $\bvec b$ represents the mean position in phase space and $a$ is a complex parameter connected with the width of the wave packet. In Ref.~\cite{Feldmeier2000} it was shown that for a hundred periodically positioned wave packets, Eqs.~\eqref{eq:finite:operators} reproduce the momentum and spatial densities intrinsic to one-dimensional Fermi systems. In Fig.~\ref{fig:1ddist}, we show that our method also reproduces this result by placing a single particle in a unit cell of size $\ell$. It is clearly visible that with increasing $\sqrt{a}/\ell$ and thus with increasing overlap of neighboring states and hence increasing influence of the antisymmetrization, the fermionic behavior becomes apparent. For $\sqrt{a}/\ell=1$, the momentum distribution does not show a sharp cutoff near the Fermi momentum because the periodic set of Gaussian wave packets does not constitute a complete set. For $\sqrt{a}/\ell\rightarrow\infty$ , the antisymmetry operator projects the Gaussian wave packets on plane waves and a sharp cutoff is reached.
%% %---------------------------------------------------------------------
%% \begin{figure}[H!]
%% \begin{center}
%% \includegraphics[width=0.45\textwidth]{density}
%% \caption{The spatial density (panels (a) and (c)) and momentum density
%%   (panels (b) and (d)) of a Gaussian wave packet with one-dimensional
%%   periodic boundary conditions within a unit cell of size $\ell$, for
%%   two different values of the width parameter $a$. For small overlap
%%   (top), the system behaves as one consisting of distinguishable
%%   particles while for large overlap (bottom) the fermionic behavior
%%   becomes manifest. To compare, the spatial distribution of a free
%%   Fermi gas is presented in panel (c) (dashed line) and the momentum
%%   distribution of the single Gaussian wave packet is presented in
%%   panel (d) (dotted line).}\label{fig:1ddist}
%% \end{center}
%% \end{figure}
%% %---------------------------------------------------------------------
Even though this one-dimensional example clearly hints at free-fermion behavior, it has to be assured that fermion-like properties of the simulations are not an artifact of the symmetry imposed by the lattice. The latter is, however, the case and it can be shown that the momentum distribution evolves toward the first Brillouin zone of the lattice \cite{vantournhout2010}. This is demonstrated in Fig.~\ref{fig:2dhexdist}, where a single wave packet is placed on a hexagonal lattice. With increasing overlap, the momentum distribution reflects the first Brillouin zone of the hexagonal lattice. This effect can be expected as the width of a single wave packet becomes much larger than the size of its unit cell, and the system behaves as plane waves in a periodic structure.
%---------------------------------------------------------------------
%---------------------------------------------------------------------

\begin{figure}[t]
\begin{center}
\includegraphics[width=0.5\textwidth]{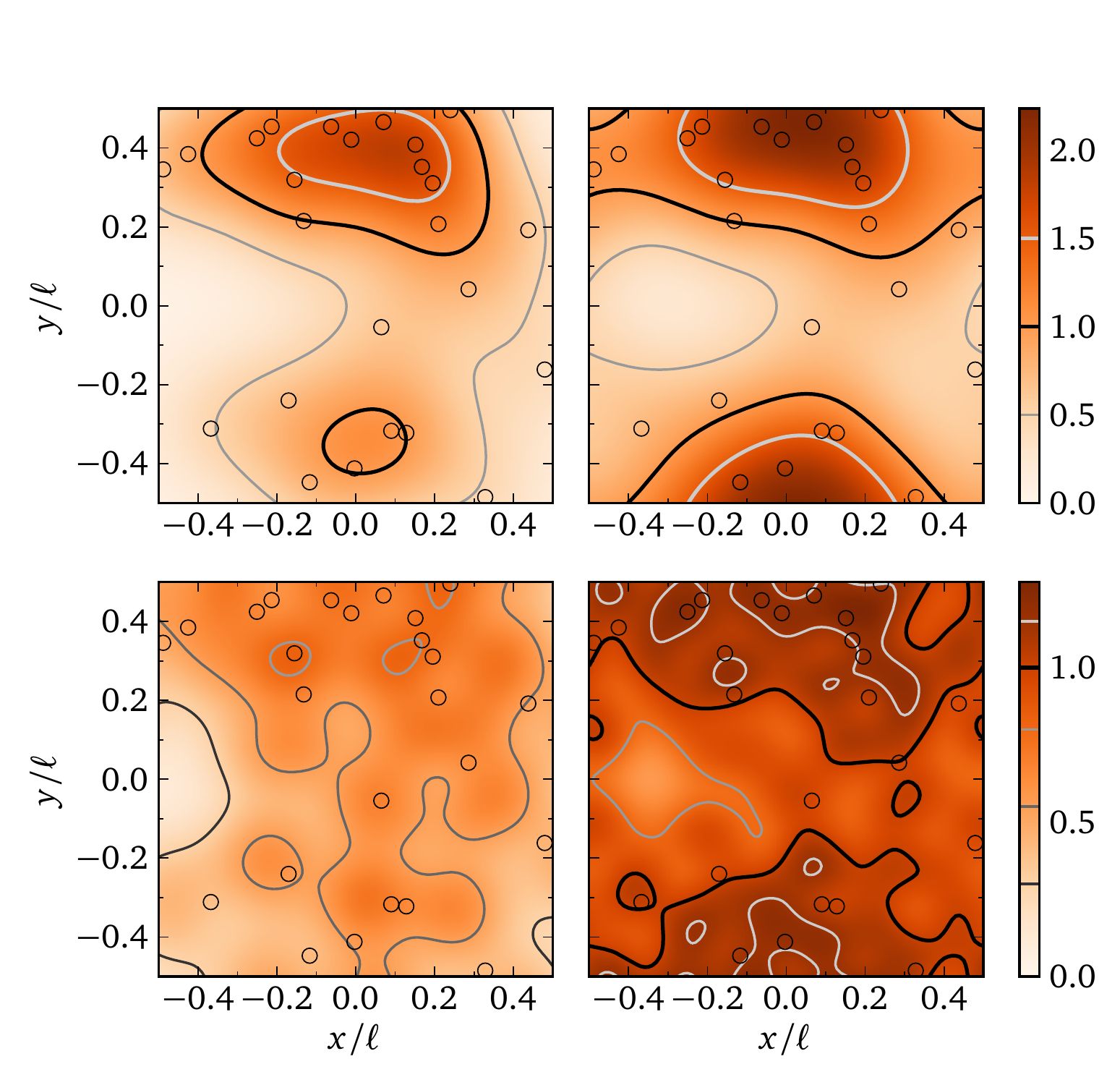}
\caption{(Color online) Comparison of the scaled spatial density distributions $\rho_x \ell^2/N$ of $N=25$ Gaussian wave packets under different conditions. The wave packets are randomly placed in the unit cell of a square lattice with unit length $\ell$ and have a variance given by $\sqrt{a}=0.2\ell$. The mean momentum of the individual wave packets is set to zero. The densities are normalized and coordinates are expressed in units of $\ell$. The upper-two panels depict distinguishable particles. The lower-two panels represent the antisymmetrised case. The left and right panels show the effect of a simulation without and with periodicity introduced, respectively, as described in the text. The black circles show the positions of the centroids of the wave packets. Note the difference in color scaling.} \label{fig:2dsqrspatial}
\end{center}
\end{figure}

\begin{figure}[t]
\begin{center}
\includegraphics[width=0.5\textwidth]{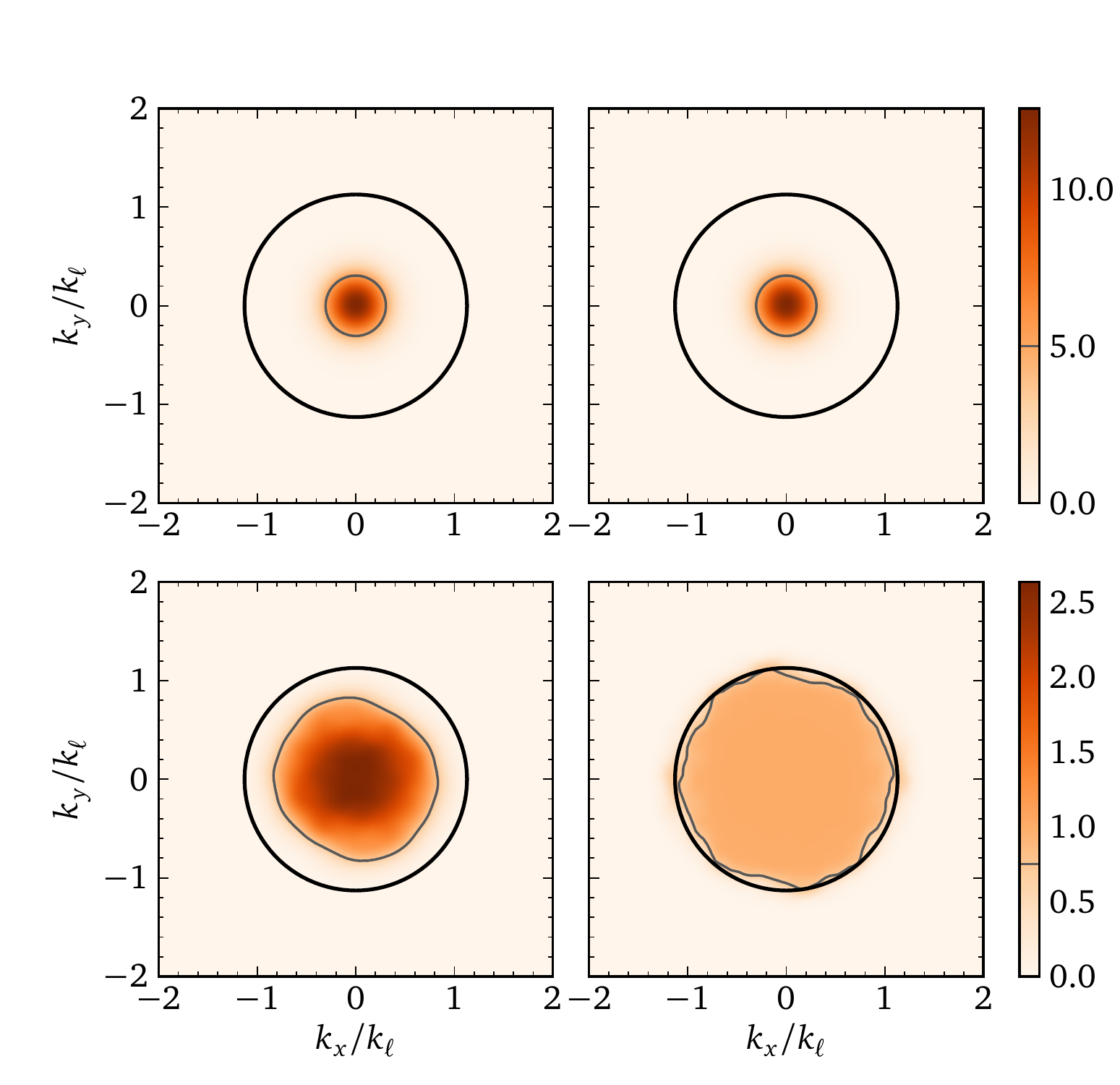}
\caption{(Color online) Comparison of the scaled momentum density distributions $\rho_k k_\ell^2/N$ of $N=25$ Gaussian wave packets under different conditions. The wave packets are randomly placed in the unit cell of a square lattice with unit length $\ell$ and have a variance given by $\sqrt{a}=0.2\ell$. The mean momentum of the individual wave packets is set to zero. The densities are normalized and coordinates are expressed in units of $k_\ell = \sqrt{N}\pi/\ell$.  The upper-two panels depict distinguishable particles. The lower-two panels represent the antisymmetrised case. The left and right panels show the effect of a simulation without and with periodicity, respectively. The black circle represents the Fermi ``sphere'' of a system of free particles with the same mean density, $k_F = \sqrt{4\pi N}/\ell$. Note the difference in color scaling.} \label{fig:2dsqrmomentum}
\end{center}
\end{figure}

As we introduced periodic boundary conditions to eliminate surface effects, a successful investigation of bulk matter obviously requires that influences of the geometry of the chosen boundary conditions are negligible. In Figs.~\ref{fig:2dsqrspatial}
%---------------------------------------------------------------------
%---------------------------------------------------------------------
and \ref{fig:2dsqrmomentum}
%---------------------------------------------------------------------
%---------------------------------------------------------------------
we investigate the effect of periodicity and antisymmetry on the spatial and momentum distributions for a situation where 25 single-particle states are randomly distributed in a square unit cell. The figures demonstrate that the spatial and momentum distribution of a periodic antisymmetrised system (lower right panels) reflect that of a free fermion system. The perturbations on the Fermi-sphere in momentum space, as well as those on the uniform distribution in coordinate space can be understood as a result of the local clustering. Hence, a simulation with the proposed periodic structure reproduces the intrinsic bulk fermion behavior, independent of the imposed periodic structure. When comparing the density distributions of the distinguishable particles with the indistinguishable fermions, it is evident that the distributions are quite different. This clearly shows that antisymmetry introduces Fermi motion, redistributing the densities. The effect of Pauli repulsion becomes visible when comparing the maxima of the density distribution with the location of the centroids in the non-periodic antisymmetrised case (lower right panel of Fig.~\ref{fig:2dsqrspatial}). When enabling periodicity, the effect is small for the distinguishable particles, but crucial for a good description of the bulk fermion system.  This clearly shows that antisymmetry, a long-range many-body correlation, should not be ignored or truncated for the bulk description. 

In summary, we propose a technique that allows one to simulate fully antisymmetrized inhomogeneous bulk fermion matter by means of localized single-particle states. A doubly-periodic structure imposed on the many-body fermion wave function replaces the classical idea of periodic boundary conditions, treating each single-particle state as an individual state, but with a truncated antisymmetry. This leads to a circulant structure in the matrix formalism for the evaluation of expectation values. This circulant structure is exploited to reduce the numerical cost of the simulation. An important merit of the proposed technique is that it becomes possible to perform antisymmetrized calculations in the limit of infinite volumes. The structure of the resulting equations resembles those of a finite fermion system, but requires an extra integration over the first Brillouin zone, reflecting the periodic boundary conditions. Thus, the truncation in the antisymmetry vanishes, yielding a formalism that allows one to compute expectation values of various operators for bulk fermion matter in a computationally feasible fashion. Although the equations only address a finite number of particles in a unit cell, they keep track of the Fermi statistics of the infinite system. We evaluate the validity of this description with a study of the spatial and momentum distribution of various lattice systems \cite{vantournhout2010} and show that our simulations convincingly reproduce free fermion behavior. This shows that full antisymmetrization can be imposed on the bulk system, a feature that has been considered hard to reach. Furthermore, the use of localized states makes the technique suitable to study inhomogeneous bulk fermion matter.

K.V.~would like to thank H.~Feldmeier and T.~Neff for enlightening discussions. This work was supported by the Fund for Scientific Research Flanders (FWO) and the Research Council of Ghent University.


\begin{thebibliography}{24}%
\bibitem{Altshuler1980}
  B.~Altshuler, A.~G.~Aronov and P.~A.~Lee, Phys. Rev. Lett., \textbf{44}, 1288 (1980)
\bibitem{Sanpera2004}
  A.~Sanpera, A.~Kantian, L.~Sanchez-Palencia, J.~Zakrzewski and M.~Lewenstein, Phys. Rev. Lett., \textbf{93}, 040401 (2004)
\bibitem{Watanabe2005}
  G.~Watanabe, T.~Maruyama, K.~Sato, K.~Yasuoka and T.~Ebisuzaki, Phys. Rev. Lett., \textbf{94}, 031101 (2005)
\bibitem{Allen1991}
  M.~P.~Allen and D.~J.~Tildesley, \emph{Computer Simulations of Liquids} (Clarendon, Oxford, 1991)
\bibitem{Aichelin1986}
  J.~Aichelin and H.~St{\"o}cker, Phys. Lett. B, \textbf{176}, 14, (1986)
\bibitem{Maruyama1996}
  T.Maruyama, K.~Niita and A.~Iwamoto, Phys. Rev. C, \textbf{53}, 297 (1996)
\bibitem{Feldmeier2000}
  H.~Feldmeier and J.~Schnack, Rev. Mod. Phys., \textbf{72}, 655 (2000)
\bibitem{Ono2004}
  A.~Ono and H.~Horiuchi, Prog. Part. Nucl. Phys., \textbf{53}, 501 (2004)
\bibitem{Slater1929} 
  J.~Slater, Phys. Rev., \textbf{34}, 1293 (1929)
\bibitem{Floquet1883}
  G.~Floquet, Ann. \'Ecole Norm. Sup., \textbf{2}, 47 (1883)
\bibitem{Bloch1929}
  F.~Bloch, Z.~Phys., \textbf{52}, 555 (1929)
\bibitem{Ashcroft1976}
  N.~W.~Ashcroft and N.~D.~Mermin, \emph{Solid State Physics} (Saunders College, 1976)
\bibitem{TinkaGammel1993}
  J.~Tinka Gammel, D.~K.~Campbell and E.~Y.~J.~Loh, Synth. Metals, \textbf{57}, 4437 (1993)
\bibitem{Lin2001}
  C.~Lin, F.~Zong and D.~Ceperley, Phys. Rev. E, \textbf{64}, 016702 (2001)
\bibitem{Wannier1937}
  G.~H.~Wannier, Phys. Rev., \textbf{52}, 191 (1937)
\bibitem{Blount1962}
  E.~J.~Blount, Solid State Physics, \textbf{13}, 305 (1962)
\bibitem{Born1912}
  M.~Born and T.~von K\'arm\'an, Physik. Z., \textbf{13}, 297 (1912)
\bibitem{vantournhout2010}
K.~Vantournhout, et al. (unpublished); 
K.~Vantournhout, \emph{Nuclear pasta with a touch of quantum}, PhD Thesis (unpublished), Ghent University (2009)
\bibitem{Baldereschi1973}
  A.~Baldereschi, Phys. Rev. B, \textbf{7}, 5212 (1973)
\bibitem{Monkhorst1976}
  H.~J.~Monkhorst and J.~D.~Pack, Phys. Rev. B, \textbf{13}, 5188 (1976)
\end{thebibliography}
\end{document}